\documentstyle[12pt]{article}

\title{Negative-energy spinors and the Fock space of
       lattice fermions at finite chemical potential
\\[1cm] (revised) \\[2cm]
}
\author{E. Mendel and L. Polley \\[1cm]
FB Physik, Oldenburg University, 26111 Oldenburg, Germany
\\[2cm]}

\begin{document}

\maketitle
\vspace{2cm}

\abstract{
Recently it was suggested that the problem of
species doubling with Kogut-Susskind lattice fermions
entails, at finite chemical potential, a
confusion of particles with antiparticles.
What happens instead is that the familiar
correspondence of positive-energy spinors to particles, and
of negative-energy spinors to antiparticles, ceases to
hold for the Kogut-Susskind time derivative.
To show this we highlight the role of the spinorial
``energy'' in the Osterwalder-Schrader
reconstruction of the Fock space of non-interacting
lattice fermions at zero temperature and nonzero chemical
potential. We consider Kogut-Susskind fermions and,
for comparison, fermions with an asymmetric one-step time
derivative.
}
\thispagestyle{empty}
\newpage

\section{Introduction}

The implementation of finite baryon density on the lattice
\cite{Kog83,HaK83} has proven to be a hard task.
Several
proposals have been made to explain the unexpectedly low
chemical potential $\mu$ at which the baryon density and
other quantities start to grow. The onset at a $\mu$ much
lower than the value $m_{\rm N}/3$, as expected for a free
nucleon gas, could be due to the enhanced number of
flavours (when using Kogut-Susskind fermions) producing a
much stronger nuclear binding than in nature
\cite{Men93,Men92}.

As the onset is so early that its energy can be almost
confounded with the existence of a `baryonic' Goldstone
mode, it was also interesting to search for possible
enhancement of mesonic modes due to some kind of lattice
artifact. In \cite{Sta93} it was pointed out that by
selecting only the positive-energy poles of the free
Kogut-Susskind propagator one still obtains energies and
baryon number densities composed of particles as well as
antiparticles. This is in contrast to Dirac fermions in
space-time continuum where positive-energy spinors always
correspond to particles and negative-energy spinors to
antiparticles. As observed in \cite{Sta93}, that
feature of continuous fermions is reproduced on the lattice
if an asymmetric, one-step time derivative is used.
(In fact, such fermions have the same Hilbert space as
Susskind's fermions \cite{Sus77} on the Hamiltonian
lattice.)
It was then suggested
in \cite{Sta93} that the Kogut-Susskind action does not
allow a clear distinction of particles and antiparticles.
That conclusion, however, is not justified.
In fact, particles and antiparticles can be reliably
identified only from the Hamiltonian and baryon
number operators in Fock space which we
therefore reconstruct below.
Apart from the usual doubling of flavours, nothing peculiar
will be seen here to occur with Kogut-Susskind fermions.
Rather, it will be the correspondence of
positive/negative-energy spinors to particles/antiparticles
which is destroyed by the Kogut-Susskind
two-step time derivative.

Intuitively, in this respect,
the essence of the technical, field-theoretic
arguments presented in the next sections is as follows.
In Euclidean time $\tau$, ``particle'' solutions of the
Dirac equation are those which remain bounded for
$\tau\rightarrow+\infty$ while ``antiparticles'' remain
bounded for $\tau\rightarrow-\infty$; this corresponds to
the Minkowskian notion of particles moving forward and
antiparticles backward in time. The rule is implicit in the
Osterwalder-Schrader reconstruction algorithm
(cf.\ remark following eqn.\ (\ref{KinvKS})).
For the Dirac equation in
continuous Euclidean time, $\dot{\psi}=-H\psi$, this
implies that the positive eigenvalues of $H$ are associated
with particles and the negative eigenvalues with
antiparticles. Now, if time is discretized, the Dirac
equation is converted into a recursion relation.
The most straightforward, asymmetric discretization would
give (with $a_{\rm T}$ being the lattice spacing and the
spinorial Hamiltonian diagonalized)
$$
\psi_{\tau+1}-\psi_{\tau} = -a_{\rm T} E ~ \psi_{\tau}
$$
Depending on the size of $ a_{\rm T} E_{\rm cutoff} $
this will distort the continuous-time dependencies somewhat
but it will preserve the qualitative features.
In contrast, the symmetric discretization would give the
recursion relation
$$
\psi_{\tau+1}-\psi_{\tau-1} = -2 a_{\rm T} E ~ \psi_{\tau}
$$
which is for {\em any} $a_{\rm T}$ a qualitatively different
type of equation. In fact, looking for ``particle''
solutions with an exponential decay $e^{-\omega\tau}$ in
the positive time direction, we now find two of them:
$$
\left. \begin{array}{r}e^{-\omega\tau} \\
           (-1)^{\tau} e^{-\omega\tau} \end{array}\right\}
\mbox{ emerging from }
\left\{ \begin{array}{c}
   a_{\rm T} E = + \sinh \omega > 0 \\
   a_{\rm T} E = - \sinh \omega < 0 \end{array}
                                                    \right.
$$
A similar degeneracy, involving both signs of spinorial
energy, is obtained with solutions bounded for
$\tau\rightarrow-\infty$.
Hence, positive spinorial energies are no longer reserved
for particles, but occur as antiparticle solutions as well.
Of course, those arguments require formal substantiation.
In the field theory that we consider here,
particles and antiparticles will be
clearly identified by the baryon number.

The Hilbert spaces corresponding to space-time lattice
fermions have all been reconstructed long ago as part of
the issue of the positivity of the transfer matrices.
This was done in \cite{Crz77,Lue77,OSe78} for Wilson
fermions, and in \cite{Sha81} for staggered fermions.
In all those
references, a direct comparison was made between matrix
elements in a standard fermion Fock space and the
Grassmann functional integral.
While this method enables
one to deal with interacting fermions from the
outset, it does not assign a Fock-space interpretation to
the path-integral variables themselves.
For Kogut-Susskind fermions, in particular, auxiliary
Grassmann integrations had to be introduced, preventing
the Fock space interpretation of the original variables. In
the present context, where we want to identify the role of
positive- and
negative-energy spinors (the Fourier transforms of the
path-integral variables) in the fermion Fock space,
we therefore prefer to work our way
through the Osterwalder-Schrader reconstruction scheme
\cite{OS73a,OS73b} which does not anticipate any knowledge
of the Hilbert space. For a
detailed introduction to this formalism
we refer the reader to \cite{Sei82}; we shall only need the
simplest of the formal concepts, and the mathematical tools
required in the following will not go beyond some suitably
organized algebra of Grassmann derivatives.

Fortunately, the
present issue allows us to restrict ourselves to
non-interacting fermions at zero temperature, but at finite
chemical potential $\mu$.

 Furthermore, we will restrict our examples to
cases where the inverse fermion matrix can be given
explicitly, in terms of spinorial energy eigenstates
and eigenvalues. Thus we consider fermion actions of the
general form
\begin{equation} \label{D+H}
S = \bar{\psi}_n K_{nn'} \psi_{n'}
\qquad n = (n_1,n_2,n_3,n_4) = (\vec{n},n_4)
\end{equation} $$
\mbox{where} \qquad K_{nn'} =
\delta_{\vec{n},\vec{n}'} D(\mu)_{n_4,n'_4}
 +
\delta_{n_4,n_4'} \frac{a_{\rm T}}{a_{\rm S}}
E_{\vec{n},\vec{n}'}
$$
such that the time-changing (derivative) part $D$ commutes
with the equal-time, spinorial ``energy'' part $E$.
This form can be achieved both for Kogut-Susskind and for
time-asymmetric \cite{Sta93} fermions. For Wilson
fermions, such a simplification does not seem to be
possible.

For the asymmetric fermions it will be advantageous to
allow for an anisotropy of the space-time lattice; hence
the appearance in (\ref{D+H}) of the spatial and temporal
lattice spacings, $a_{\rm S}$ and $a_{\rm T}$.

In section 2 we recall the basic steps in the procedure of
Osterwalder-Schrader reconstruction, and the standard tools
we have at hand for non-interacting fermions. In section 3
we apply this, in some detail, to fermions with a one-step
time derivative. We thus recover the familiar
zero-temperature scenario at finite chemical potential.
In section 4 we provide the analogous expressions for
Kogut-Susskind fermions. A comparison of
the operator expressions for the energy and for the baryon
number in Fock space (as opposed to the linear space
spanned by the Grassmann variables) will show that a
chemical potential enhancing the baryons will deplete the
antibaryons.
In the concluding section 5 we discuss relations with the
positivity of the transfer matrices and with finite
temperature.

\section{Basic ideas of reconstructing $\hat{H}$ and
$\hat{B}$}

{}From the path-integal expectation values and correlations
we wish to recover the multi-fermion Hilbert space, the
scalar product of the Hilbert space vectors, and the
operators of time evolution and baryon number.
The basic ideas are as follows \cite{Sei82}.
Both bra and ket vectors contribute
to $\langle\cdots\rangle_{\rm path~integral}$. The ket
vectors are the functionals of Grassmann variables
$\psi_{\tau}$ and $\bar{\psi}_{\tau}$ with Euclidean times
$\tau>0$ (more precisely, they are equivalence classes of
such functionals; cf.\ below). The bra vectors are the same
with $\tau<0$. Thus bra and ket functionals
are related through a time-reversal operation $\Theta$.
On Grassmann variables it acts as
$$
 \Theta \psi(\tau) = \bar{\psi}(-\tau)   \qquad
 \Theta \bar{\psi}(\tau) = \psi(-\tau)
$$
and is extended as an antilinear operation on functionals
of $\psi$ and $\bar{\psi}$. We shall adopt the rule that
complex conjugation of a product of Grassmann variables
implies reversed ordering of factors.

The scalar product of a bra and ket vector is given by a
path-integral correlation:
if $F$ and $G$ are ket functionals, and $|F\rangle$ and
$|G\rangle$ their interpretations as Hilbert space vectors,
their scalar product is
$$
\langle F | G \rangle  =
\langle \Theta F ~ G \rangle _{\rm path~integral} =
 \frac1Z
\int  [{\rm d}\psi] [{\rm d}\bar{\psi}] ~ e^S ~
\Theta F [ \bar{\psi},\psi]_{\tau<0} ~
       G [ \bar{\psi},\psi]_{\tau>0}
$$
For a quantum-statistical interpretation the
scalar product must satisfy $\langle F | F \rangle \geq 0$
which it does in all the cases we consider.
Null-norm functionals
(which have $\langle F|F \rangle=0$ although $F\neq0$)
can then be factored away.
This will, in fact, simplify the structure of the Hilbert
space.

We will extract the transfer matrix operator $\hat{T}$
from the path integral correlation
 $\langle F|\hat{T}|G\rangle$ in such a way that it will
reproduce these matrix elements for arbitrary $F$ and $G$.
The transfer matrix, which defines the true energy
(as opposed to spinorial ``energy'')
by $\hat{T}=e^{-\hat{H}}$,
derives from a unit shift of the time
index on all the Grassmann variables in a ket functional
$$
\langle ~F[\bar{\psi}_{\tau},\psi_{\tau}] ~|~ \hat{T}~
| ~G [\bar{\psi}_{\tau},\psi_{\tau}]~ \rangle =
\langle ~F [\bar{\psi}_{\tau},\psi_{\tau}] ~| ~
G[\bar{\psi}_{\tau+1},\psi_{\tau+1}] ~\rangle
$$
The baryon number operator will likewise
be identified from its
matrix elements $\langle F|\hat{B}|G\rangle$, i.e.\ from
the path integral correlation $\langle\Theta F~B~G\rangle$
in which $B[\bar{\psi},\psi] = \partial S / \partial \mu $
is the usual observable.

It will be essential in the following to have an explicit
representation of Grass\-mann-in\-te\-g\-ral
expectation values and
correlations at hand. In the most general finite-dimensional
case, if $\psi_n$ and $\bar{\psi}_n$ are the integration
variables and if $O[\psi,\bar{\psi}]$ is any observable or
correlation in terms of them, then \cite{Sak85}
$$
\langle O \rangle_{\rm path~integral} = \frac1{\det K}
\int \prod_n {\rm d}\psi_n {\rm d}\bar{\psi}_n \exp \left(
\sum_{n,m} \bar{\psi}_n K_{nm} \psi_m \right)
O[\psi,\bar{\psi}]
$$
\begin{equation} \label{expc}
= \left. \exp \left( \sum_{n,m} K^{-1}_{nm}
   \frac{\partial}{\partial \psi_n}
\frac{\partial} {\partial \bar{\psi}_m} \right)
O[\psi,\bar{\psi}] \right|_{\psi=\bar{\psi}=0}
\end{equation}
To keep the notation as simple as possible, we shall refer
only to Grassmann derivatives ``from left to right''
($\partial/\partial\psi=\stackrel{\rightarrow}{\partial}$).
The reader who wishes to check the crucial signs for
himself will find it convenient to use
Grassmann derivatives ``from right to left'' as well,
noting that
$  \stackrel{\rightarrow}{\partial}
   \stackrel{\rightarrow}{\partial} \bar{F}\bar{G} =
 -  \bar{F} \stackrel{\leftarrow}{\partial}
      \stackrel{\rightarrow}{\partial} G           $

We shall be only concerned here with the Euclidean time
structure of the inverse action kernel $K^{-1}_{nm}$, where
$n=(n_1,n_2,n_3, n_4)$ etc.
We shall use two different labelings of time slices:
$n_4=0,\pm1,\pm2,\ldots$ as appropriate for the space-time
functional-integral approach, and $\tau=\pm1, \pm2, \ldots$
which will emphasize later on the correspondence between
bra
and ket functionals. For later convenience we note that
$$
\tau = n_4 \quad\mbox{for}\quad n_4>0 \qquad \tau=n_4-1
           \quad\mbox{for}\quad n_4\leq0
$$
With respect to the spatial indices, we transform into a
basis in which the equal-time part of the action is
diagonal. For Kogut-Susskind fermions
this involves spatial Fourier transformation and linear
combination of the momenta at the edges of the Brillouin
zone. Let $p$ denote a suitable (multi)index
for the basis vectors. For example, $p$ can be chosen as
\begin{equation} \label{multiindex}
p = (p_1,p_2,p_3,\sigma,s,f) \qquad
p_k\in\left]-\frac{\pi}{2a_{\rm S}}
,\frac{\pi}{2a_{\rm S}}\right],
\quad\sigma=\pm 1,\quad s=\pm\frac12,\quad f=u,d
\end{equation}
where $p_k$ are the components of spatial momentum.
The rest of the Brillouin zone, reachable by addition of
``edge'' momenta $(\pm\pi/a,\pm\pi/a,\pm\pi/a)$
corresponds to the degrees of freedom of spinorial
``energy'' sign $\sigma$, spin $s$, and flavour $f$
\cite{Smi84}. In the following,
$\sigma(p)$ will be of particular interest to us. Using
$$
    E(p) = \sqrt{\sum_k \sin^2 p_k + m^2}
$$
the spinorial ``energy'' is $\sigma(p) E(p)$.

We can now rewrite the action kernel of
equation (\ref{D+H}) as
$$
K_{nn'} = \delta_{p,p'} \left( D(\mu)_{n_4,n'_4}
  + \delta_{n_4,n'_4} \frac{a_{\rm T}}{a_{\rm S}}
                 \sigma(p) E(p)
                                                  \right)
$$
In the following we reconstruct the fermion Fock space,
and the operators of energy and baryon number in particular,
for two versions of the time derivative.

\section{Asymmetric time derivative}

We first analyse fermions with a one-step, asymmetric time
derivative. This will lead to expectation values
(and operators) for the
energy and baryon number resembling those familiar from
space-time continuum \cite{Sta93}.

The time-asymmetric action kernel, corresponding to equation
(2.15) of ref.\ \cite{Sta93}, is
$$
K_{nn'} = \frac{a_{\rm T}}{a_{\rm S}} m \delta_{nn'}
+ \frac12\frac{a_{\rm T}}{a_{\rm S}} \sum_{k=1}^3
 \gamma_k~(\delta_{n,n'+\hat{k}}-\delta_{n,n'-\hat{k}})
+ \gamma_4~(e^{\mu}\delta_{n,n'+\hat{4}}-\delta_{n,n'})
$$
where the Euclidean gamma matrices are characterized by
$\{\gamma_{\nu},\gamma_{\nu'}\}=\delta_{\nu\nu'}$ and
$\gamma_{\nu}^{\dag}=\gamma_{\nu}$. To reduce the fermion
degeneracy, the $4\times4$ matrix structure can be
eliminated by first absorbing a $\gamma_4$ factor into the
$\bar{\psi}$ integration variables and then diagonalizing
in spin and restricting to one of the components
by analogy to Kawamoto-Smit.
Hence let us define $\alpha_k=i\gamma_4\gamma_k$ and
$$
\bar{\chi}(n) = \bar{\psi}(n) ~ \gamma_4 ~
 \alpha_3^{n_3} \alpha_2^{n_2} \alpha_1^{n_1}     \qquad
\chi(n)=\alpha_1^{n_1}\alpha_2^{n_2}\alpha_3^{n_3}\psi(n)
$$
With respect to the $\chi$ and $\bar{\chi}$ spinors the
action kernel is
$$
K_{nn'} =
  \frac{a_{\rm T}}{a_{\rm S}} m \delta_{nn'}
                                      \gamma_4 \Gamma_4(n)
+ \frac12\frac{a_{\rm T}}{a_{\rm S}} \sum_{k=1}^3
 \Gamma_k(n)~(\delta_{n,n'+\hat{k}}-\delta_{n,n'-\hat{k}})
+ e^{\mu}\delta_{n,n'+\hat{4}}-\delta_{n,n'}
$$
where $\Gamma_{\nu}(n)=(-1)^{n_1+\cdots+n_{\nu-1}}$.
A $\gamma_4$ matrix has remained in the mass term but,
assuming it to be diagonalized,
the spinor components are now decoupled
so that we can build a flavour-reduced fermionic system
by projecting onto the first component. We note that the
equal-time part of this action coincides with Susskind's
spatial fermions \cite{Sus77}.
After spatial Fourier transformation the asymmetric kernel
becomes
$$
K_{nn'} = \delta_{pp'} \left( ( e^{\mu} \delta_{n_4+1,n_4'}
        - \delta_{ n_4, n_4'} )
+\sigma(p) E(p) \frac{a_{\rm T}}{a_{\rm S}}
\delta_{n_4,n'_4} \right)
$$
To evaluate any scalar product or matrix element,
such as $\langle (\Theta F) \,B\,G\rangle$, using
(\ref{expc}), we need to know the inverse of the action
kernel.
Using
$$
\epsilon \equiv \ln \left(
 1-\frac{a_{\rm T}}{a_{\rm S}} \sigma(p) E(p)\right)
$$
we obtain
\begin{equation} \label{Kinv1}
K^{-1}_{nn'} = \left\{ \begin{array}{cl}
   - e^{-\epsilon} e^{(\epsilon-\mu)( n_4- n_4')} &
                            \mbox{if }  n_4\leq n_4' \\
         0  & \mbox{if }  n_4 >  n_4'
      \end{array}\right\} \delta_{pp'}
        \quad\mbox{for } \epsilon > \mu
\end{equation} \begin{equation}  \label{Kinv2}
K^{-1}_{nn'} = \left\{ \begin{array}{cl}
 0  & \mbox{if }  n_4 \leq  n_4' \\
   e^{-\epsilon} e^{(\epsilon-\mu)( n_4- n_4')} &
                            \mbox{if }  n_4 >  n_4'
      \end{array}\right\} \delta_{pp'}
        \quad\mbox{for } \epsilon < \mu
\end{equation}
It should be mentioned here that the zeros above will
produce zero-norm states later on.
Right at $\epsilon=\mu$ the inverse kernel is unambiguously
determined by the antiperiodic boundary condition with
respect to time. The appropriate limit is
$$
K^{-1} = \frac12 \left( K^{-1}_{\epsilon\rightarrow\mu-0} +
                 K^{-1}_{\epsilon\rightarrow\mu+0} \right)
$$
It is
convenient to adapt operation (\ref{expc}) to the special
case of $O=\bar{F}\cdot G$ with $\bar{F}$ a negative-time
and $G$ a positive-time functional. (This also covers the
case of a matrix element like $\langle\Theta F
\,B\,G\rangle$.)
Sorting out the derivatives acting on $\bar{F}$ or $G$
exclusively, we can write
\begin{equation} \label{GGG}
\exp \sum_{nn'} K^{-1}_{nn'}
\frac{\partial}{\partial\psi_n}
\frac{\partial}{\partial\bar{\psi}_{n'}} ~~ \bar{F} G =
{\cal C}_0 \left( ({\cal C}_-\bar{F}) ({\cal C}_+ G) \right)
\end{equation}
where the ${\cal C}$ stand for ``contractions'' and
$$
{\cal C}_{\pm} = \exp \sum_p
    \sum_{\begin{array}{c}  \tau,\tau'> 0
\\                          \tau,\tau'< 0 \end{array} }
K^{-1}_{\tau\tau'}(p) \frac{\partial}{\partial\psi_{p\tau}}
\frac{\partial}{\partial\bar{\psi}_{p\tau'}}
$$
while the remaining ``mixing'' term is
\begin{equation} \label{G0as}
{\cal C}_0 = \exp \sum_{p:~\epsilon>\mu} (-e^{-\mu})
\left( \sum_{\tau>0} e^{(\mu-\epsilon)\tau}
 \frac{\partial}{\partial\psi_{p,-\tau}} \right)
\left( \sum_{\tau'>0} e^{(\mu-\epsilon)\tau'}
\frac{\partial}{\partial\bar{\psi}_{p,\tau'}} \right) \times
\end{equation}
$$ \times
\exp \sum_{p:~\epsilon<\mu} e^{\mu-2\epsilon}
\left( \sum_{\tau>0} e^{(\epsilon-\mu)\tau}
 \frac{\partial}{\partial\psi_{p,\tau}} \right)
\left( \sum_{\tau'>0} e^{(\epsilon-\mu)\tau'}
\frac{\partial}{\partial\bar{\psi}_{p,-\tau'}} \right)
$$
The operations ${\cal C}_+$ and ${\cal C}_-$ only modify
the ket and bra functionals, respectively.
The operation intertwining kets and bras is ${\cal C}_0$,
and the last step is to discard all remaining $\psi$s and
$\bar{\psi}$s. It will be most convenient to formulate the
fermion Hilbert space in terms of $\bar{F}'={\cal C}_-
\bar{F}$ and $G'={\cal C}_+ G$, i.e., at the ${\cal C}_0$
stage after the ${\cal C}_+$ and ${\cal C}_-$ operations.

We are aiming at operator expressions for the energy
(transfer matrix $\hat{T}$ or Hamiltonian $\hat{H}$) and
for the baryon number. As ${\cal C}_+$ and ${\cal C}_-$
depend on $\tau-\tau'$ only, they commute with the time
evolution. Thus $\hat{T}$ can be determined at the ${\cal
C}_0$ stage immediately; see below. The baryon number
operator is most easily identified from the path-integral
observable ``at $\tau=0$'' (i.e., involving $\tau=-1$ and
$\tau=1$) which is
\begin{equation} \label{Bas}
B = \frac{\partial S}{\partial\mu} = \sum_{\vec{n}} e^{\mu}
\bar{\psi}_{\vec{n},-1} \psi_{\vec{n},1} = \sum_p
e^{\mu} \bar{\psi}_{p,-1} \psi_{p,1}
\end{equation}
In the second equality we changed to the eigenbasis of
the spinorial energy, and arranged Grassmann variables in
time order.
$\bar{\psi}_{p,-1}$ and $\psi_{p,1}$ can now be
interpreted as multiplication operators acting on bra and
ket functionals, respectively. Thus the matrix elements of
the baryon number operator are
$$
\langle F | \hat{B} | G \rangle = e^{\mu}
\langle\Theta(\psi_{p,1}F)~\psi_{p,1} G
\rangle_{\rm path~integral} = e^{\mu} {\cal C}_0
\left\{ {\cal C}_-(\Theta\psi_{p,1}F)
{\cal C}_+\psi_{p,1} G \right\}
$$
To bring the field operators (multiplication operators) of
the above equation to the ${\cal C}_0$ stage, we
apply the Baker-Hausdorff formula, obtaining for example
the following primed form (with any ket functional $G$)
\begin{equation} \label{a+adag}
{\cal C}_+ \psi_{p,1} G = \left( \psi_{p,1} + \left\{
\begin{array}{cc} 1 & \epsilon>\mu \\ 0 & \epsilon<\mu
\end{array} \right\} \times e^{-\mu}
\sum_{\tau>0} e^{(\mu-\epsilon)\tau}
\frac{\partial}{\partial\bar{\psi}_{p,\tau}} \right) G'
\end{equation}
On the bra side, we obtain the $\Theta$-reflected
expression. In a similar way, by commuting $\psi$s and
$\bar{\psi}$s through ${\cal C}_0$ one obtains the adjoints
of field operators. For example, we have
$$
{\cal C}_0 \bar{\psi}_{p,-1} = \left( \bar{\psi}_{p,-1}
+ \left\{
\begin{array}{cc} 0 & \epsilon>\mu \\ 1 & \epsilon<\mu
\end{array} \right\} \times e^{-\epsilon}
\sum_{\tau>0} e^{(\epsilon-\mu)\tau}
\frac{\partial}{\partial\psi_{p,\tau}} \right){\cal C}_0
$$ $$
= \bar{\psi}_{p,-1} {\cal C}_0 +  {\cal C}_0
 \left( \left\{
\begin{array}{cc} 0 & \epsilon>\mu \\ 1 & \epsilon<\mu
\end{array} \right\} \times e^{-\epsilon}
\sum_{\tau>0} e^{(\epsilon-\mu)\tau}
\frac{\partial}{\partial\psi_{p,\tau}} \right)
$$
The $\bar{\psi}_{p,-1}$ on the {\sc rhs} is discarded when
all $\psi$ and $\bar{\psi}$ are set to zero in the final
step of the scalar product. The bracketed expression on the
{\sc rhs} is the ket-adjoint to the bra-multiplication by
$\bar{\psi}_{p,-1}$. To further develop our example, this
can be used to rewrite
$
{\cal C}_0\left\{(\Theta\psi_{p,1}F')\psi_{p,1}G'\right\}
$ as $
{\cal C}_0\left\{(\Theta F')\psi_{p,1}^{\dag}\psi_{p,1}G'
\right\}
$
which is part of the baryon number operator we are
interested in. From an expression such as the last one,
in which all operator action is thrown on the kets,
one finally abstracts away the functionals and the scalar
product operation.

On the ${\cal C}_0$ stage, as we see from (\ref{G0as})
and (\ref{a+adag}), Grassmann derivatives occur only in
those linear combinations of time indices which,
for each momentum $p$, correspond to ``physical'' time
dependency. Linear combinations of $\psi_{\tau}$ and
$\bar{\psi}_{\tau}$ orthogonal to it drop out from all
matrix elements (they create the null-norm functionals).
The only relevant variables are
$\Psi(p)\propto\sum_{\tau>0} e^{(\epsilon-\mu)\tau}$ and
$\bar{\Psi}(p)\propto\sum_{\tau>0} e^{(\mu-\epsilon)\tau}$.
They are conveniently defined by
$$
\sum_{\tau>0} e^{(\mu-\epsilon)\tau}
\frac{\partial}{\partial\bar{\psi}_{p,\tau}} =:
\frac{\partial}{\partial\bar{\Psi}(p)} \qquad \epsilon>\mu
$$ $$
\sum_{\tau>0} e^{(\epsilon-\mu)\tau}
\frac{\partial}{\partial\psi_{p,\tau}} =:
\frac{\partial}{\partial\Psi(p)} \qquad \epsilon<\mu
$$
so that, in the ket functionals, we have
\begin{equation} \label{Psi} \begin{array}{c}
\bar{\psi}_{p,\tau} = e^{(\mu-\epsilon)\tau}
\bar{\Psi}(p) + {\rm orth.} \qquad \epsilon > \mu
\\
\psi_{p,\tau} = e^{(\epsilon-\mu)\tau}
\Psi(p) + {\rm orth.} \qquad \epsilon < \mu
\end{array} \end{equation}
(The $\Psi(p)$ and $\bar{\Psi}(p)$ actually represent
equivalence classes.)
Omitting the irrelevant ``orthogonal'' parts we obtain
functionals of $\Psi(p)$ and $\bar{\Psi}(p)$ only.
We note that the $\Psi$ are
non-null only for $\epsilon<\mu$ (in ket functionals) and
the $\bar{\Psi}$ only for $\epsilon>\mu$.
This will be different for Kogut-Susskind fermions, leading
to an additional doubling of species there.

In terms of the new variables, the baryon number
(\ref{Bas}) for time-asymmetric fermions takes the simple
operator form
\begin{equation} \label{hatBas}
\hat{B} = \sum_{p:\,\epsilon>\mu}
\bar{\Psi}(p) \frac{\partial}{\partial\bar{\Psi}(p)}
  + \sum_{p:\,\epsilon<\mu} \left( 1 -
\Psi(p) \frac{\partial}{\partial\Psi(p)} \right)
\end{equation}
At the ${\cal C}_0$ stage, the transfer matrix is also
easily identified. If all Grassmann variables in a ket
functional are shifted by one time step, equation
(\ref{Psi}) shows that
$\bar{\Psi}(p)$ picks up a factor of $e^{\mu-\epsilon}$, and
$\Psi(p)$ a factor of $e^{\epsilon-\mu}$. The generator of
such a transformation is
\begin{equation} \label{Has}
\hat{H} - \mu \hat{B}  = -\log \hat{T} =
\sum_{p:\,\epsilon>\mu} (\epsilon-\mu)
\bar{\Psi}(p) \frac{\partial}{\partial\bar{\Psi}(p)} +
 \sum_{p:\,\epsilon<\mu} (-\epsilon+\mu)
\Psi(p) \frac{\partial}{\partial\Psi(p)}
\end{equation}
We can rewrite this in the standard Fock-space form,
using $\epsilon_m = \log (1+\frac{a_{\rm T}}{a_{\rm S}}m)$:
\begin{equation} \label{HasFock}
\hat{H}-\mu\hat{B} = \sum_{p:\,\epsilon>\epsilon_m}
                      (\epsilon-\mu) b^{\ast}(p) b(p)
+ \sum_{p:\,\epsilon>\epsilon_m}
          (\epsilon+\mu) \bar{b}^{\ast}(p) \bar{b}(p)
+ \sum_{p:\,\mu>\epsilon>\epsilon_m} (\epsilon-\mu)
\end{equation}
\begin{equation}  \label{BasFock}
  \hat{B} = \sum_{p:\,\epsilon>\epsilon_m}
\left( b^{\ast}(p) b(p) -
  \bar{b}^{\ast}(p) \bar{b}(p) \right)
+ \sum_{p:\,\epsilon<-\epsilon_m}1
\end{equation}
where we have used
$\epsilon_m = \log (1+\frac{a_{\rm T}}{a_{\rm S}}m)$
and defined creation and annihilation operators
for antibaryons (denoting by $\bar{p}$ the multiindex $p$
as in (\ref{multiindex}) with a reversed sign of spinorial
energy) by
$$
\bar{b}^{\ast}(p) = \Psi(\bar{p})  \qquad
\bar{b}(p) = \partial/\partial\Psi(\bar{p})
\qquad \epsilon(p) > \epsilon_m
$$
and for baryons by
$$
b^{\ast}(p) = \left\{ \begin{array}{cc}
      \bar{\Psi}(p) & \epsilon>\mu \\
      \partial/\partial\Psi(p) &
                \mu>\epsilon>\epsilon_m
        \end{array} \right. \qquad
b(p) = \left\{ \begin{array}{cc}
      \partial/\partial\bar{\Psi}(p) & \epsilon>\mu \\
 \Psi(p) & \mu>\epsilon>\epsilon_m \\
        \end{array} \right.
$$
The $b^{\ast}$ and $\bar{b}^{\ast}$
satisfy canonical anticommutation relations with
$b$ and $\bar{b}$, implying standard identification for
number operators. However, for $\epsilon(p)$ or $\mu$ large
(comparable to the {\sc uv} cutoff) they deviate from the
hermitian adjoints of $b$ and $\bar{b}$ by a real factor
$$
\bar{b}^{\dag}(p) = e^{-2\epsilon-\mu}
\bar{b}^{\ast}(p) \qquad
\mbox{ for all }\epsilon>\epsilon_m
$$ $$
b^{\dag}(p) = e^{\mu}b^{\ast}(p) \quad
    \epsilon>\mu \qquad \qquad
b^{\dag}(p) = e^{\mu-2\epsilon}b^{\ast}(p)
\quad \mu>\epsilon>\epsilon_m
$$
As we see from (\ref{Has}) the functional representing the
ground state $|\mu\rangle$ at chemical potential $\mu$ is
$G_0=1$. In terms of Fock space operators it is
characterized by
\begin{eqnarray} \nonumber
\bar{b}(p) |\mu\rangle = 0 & & \mbox{all }p \\
      b(p) |\mu\rangle = 0 & & \mbox{for }\epsilon>\mu
                                               \label{vac}\\
b^{\dag}(p)|\mu\rangle = 0 & & \mbox{for }
                      \mu>\epsilon>\epsilon_m
\nonumber
\end{eqnarray}
Thus the ground state is filled with baryons of all
momenta, spins and flavours with energies in the range
$\mu>\epsilon(p)>\epsilon_m$.

The zero-point energy
in (\ref{HasFock}) is due to the above definition of
$\hat{H}-\mu\hat{B}$ as a plain generator of a temporal
translation. Had we defined the baryon number operator
as a plain generator of phase rotations, we would
have avoided the divergent but
$\mu$-independent contribution in (\ref{BasFock})
which we must now subtract by hand.

It should be noted that definition (\ref{HasFock}), based
on a single time step, is valid only if $\epsilon$ in the
above definition
 $\epsilon = \ln (1-\frac{a_{\rm T}}{a_{\rm S}}\sigma(p)
E(p))$
is a real number for all energy-momentum indices $p$.
Since the maximal eigenvalue of the free staggered
Hamiltonian is $\sqrt{3+m^2}$, $m$ being the fermion mass,
the condition is $a_{\rm T}< a_{\rm S} / \sqrt{3+m^2}$.
If the temporal spacing is too coarse to satisfy this,
then one can still define a two-step Hamiltonian,
      $\hat{H}-\mu\hat{B} = -\frac12\log\hat{T}^2$
as customary for staggered fermions.

\section{Kogut-Susskind time derivative}

We now consider the two-step time derivative as
it is used with Kogut-Susskind fermions \cite{Kog83,HaK83}.
The action kernel, allowing for a space-time anisotropy, is
$$
K_{nn'} = \frac{a_{\rm T}}{a_{\rm S}} m \delta_{nn'}
+ \frac12  \frac{a_{\rm T}}{a_{\rm S}} \sum_{k=1}^3
\Gamma_k(n) (\delta_{n+\hat{k},n'}-\delta_{n-\hat{k},n'})
+ \Gamma_4(n)(e^{ \mu}\delta_{n+\hat{4},n'}
            - e^{-\mu}\delta_{n-\hat{4},n'})
$$
Absorbing a factor of $\Gamma_4$ into the $\bar{\psi}$
integration and Fourier transforming with respect to
$\vec{n}$ we arrive at
$$
K_{n n'} = \delta_{p p'} K_{n_4 n_4'}(p)
\mbox{ where }  K_{n_4 n_4'}(p)
 = \textstyle{\frac12} \left( e^{\mu} \delta_{ n_4+1, n_4'}
  - e^{-\mu} \delta_{ n_4-1, n_4'} \right) + \sigma(p) E(p)
    \frac{a_{\rm T}}{a_{\rm S}}     \delta_{ n_4 n_4'}
$$
Let us restrict to $\mu\geq 0$ and put
$$
\epsilon \equiv
{\rm arsinh}\left(\frac{a_{\rm T}}{a_{\rm S}}\sigma E
\right)
$$
The inverse kernel $K^{-1}_{ n_4, n_4'}(p)$ then reads
$$
\frac{e^{-\mu( n_4- n_4')}
e^{-\epsilon| n_4- n_4'|}}{\cosh \epsilon} \times \left\{
\begin{array}{cl} 1 & \mbox{for } n_4\geq n_4' \\
 (-1)^{ n_4- n_4'} & \mbox{for }  n_4\leq n_4' \end{array}
\right. \qquad \epsilon > \mu
$$
\begin{equation} \label{KinvKS}
\frac{e^{-\mu( n_4- n_4')}
e^{\epsilon| n_4- n_4'|}}{\cosh \epsilon} \times \left\{
\begin{array}{cl} (-1)^{ n_4- n_4'+1} &
\mbox{for } n_4\geq n_4' \\
 -1 & \mbox{for }  n_4\leq n_4' \end{array}
\right. \qquad \epsilon < -\mu
\end{equation}
$$
e^{-\mu( n_4- n_4')}\frac{e^{-\epsilon( n_4- n_4')}
  - (-1)^{ n_4- n_4'} e^{\epsilon( n_4- n_4')}}
{\cosh\epsilon} \times
\left\{\begin{array}{cl}
             1  & \mbox{for } n_4> n_4' \\
             0  & \mbox{for } n_4\leq n_4'\end{array}
\right. \qquad |\epsilon|<\mu
$$
A general remark may be in order here. The inverse kernel
consists of positive-time and negative-time solutions of the
discretized Dirac equation, suitably patched together at
$n_4=n_4'$. In eqn.\ (\ref{expc}) the index $n_4$ always
goes with a $\partial/\partial\psi$.
If for a particular solution $n_4$ runs over
the positive half-line, then $\psi$s are detected in the
ket functional; if $n_4$ runs over the negative half-line,
the $\psi$s are in the bra functional where they are
interpreted as complex-conjugates. Whatever corresponds
to a ``particle'' in one case thus corresponds to an
``antiparticle'' in the other.

To carry on with the technicalities, we change to time
labels $\tau,\tau'$
and split the summation over time according to
(\ref{GGG}). In particular, the intertwining operation
analogous to (\ref{G0as}) now reads
$$
{\cal C}_0 =
\exp \sum_{p:~\epsilon>-\mu} \frac{e^{\epsilon+\mu}}
{\cosh\epsilon} \left( \sum_{\tau>0}
e^{-(\epsilon+\mu)\tau}
\frac{\partial}{\partial\psi_{p,\tau}} \right)
\left( \sum_{\tau'>0}
e^{-(\epsilon+\mu)\tau'}
\frac{\partial}{\partial\bar{\psi}_{p,-\tau'}} \right)
$$ $$
\times
\exp \sum_{p:~\epsilon<\mu} \frac{e^{\mu-\epsilon}}
{\cosh\epsilon} \left( \sum_{\tau>0}
(-1)^{\tau} e^{(\epsilon-\mu)\tau}
\frac{\partial}{\partial\psi_{p,\tau}} \right)
\left( \sum_{\tau'>0}
(-1)^{\tau'} e^{(\epsilon-\mu)\tau'}
\frac{\partial}{\partial\bar{\psi}_{p,-\tau'}} \right)
$$ $$
\times
\exp \sum_{p:~\epsilon>\mu} \frac{-e^{\epsilon-\mu}}
{\cosh\epsilon} \left( \sum_{\tau>0}
(-1)^{\tau} e^{(\mu-\epsilon)\tau}
\frac{\partial}{\partial\psi_{p,-\tau}} \right)
\left( \sum_{\tau'>0}
(-1)^{\tau'} e^{(\mu-\epsilon)\tau'}
\frac{\partial}{\partial\bar{\psi}_{p,\tau'}} \right)
$$ $$
\times
\exp \sum_{p:~\epsilon<-\mu} \frac{-e^{-\epsilon-\mu}}
{\cosh\epsilon} \left( \sum_{\tau>0}
e^{(\epsilon+\mu)\tau}
\frac{\partial}{\partial\psi_{p,-\tau}} \right)
\left( \sum_{\tau'>0}
e^{(\epsilon+\mu)\tau'}
\frac{\partial}{\partial\bar{\psi}_{p,\tau'}} \right)
$$
{}From this we can identify the relevant Grassmann variables
at the ${\cal C}_0$ stage of the scalar product. The
procedure is largely the same as with asymmetric fermions.
Again, the new variables can be conveniently defined by
$$
\sum_{\tau>0} e^{-|\epsilon+\mu|\tau}
\frac{\partial}{\partial\psi_{p,\tau}} =:
\frac{\partial}{\partial\Psi(p,+)}
\qquad\qquad
\sum_{\tau>0} e^{-|\epsilon+\mu|\tau}
\frac{\partial}{\partial\bar{\psi}_{p,\tau}} =:
\frac{\partial}{\partial\bar{\Psi}(p,+)}
$$ $$
\sum_{\tau>0} (-1)^{\tau} e^{-|\epsilon-\mu|\tau}
\frac{\partial}{\partial\psi_{p,\tau}} =:
\frac{\partial}{\partial\Psi(p,-)}
\qquad\qquad
\sum_{\tau>0} (-1)^{\tau} e^{-|\epsilon-\mu|\tau}
\frac{\partial}{\partial\bar{\psi}_{p,\tau}} =:
\frac{\partial}{\partial\bar{\Psi}(p,-)}
$$
so that, in the ket functionals, the Grassmann variables
decompose as follows:
\begin{eqnarray}
\epsilon>\mu &\quad
\bar{\psi}_{p,\tau} = (-1)^{\tau}
     e^{(\mu-\epsilon)\tau} \bar{\Psi}(p,-) + {\rm orth.}
&\quad
\psi_{p,\tau} = e^{-(\epsilon+\mu)\tau} \Psi(p,+)
 + {\rm orth.}
\nonumber \\
|\epsilon|<\mu &\quad
   \bar{\psi}_{p,\tau} = 0  + {\rm orth.}
 &\quad
\psi_{p,\tau} = e^{-(\epsilon+\mu)\tau}
 \Psi(p,+)
   + (-1)^{\tau} e^{(\epsilon-\mu)\tau} \Psi(p,-)
\nonumber \\
& &\quad ~~~~~~~~~~~~~~~~~~~~ + ~~{\rm orth.}\label{PsiKS}\\
\epsilon<-\mu &\quad
\bar{\psi}_{p,\tau} = e^{(\epsilon+\mu)\tau}
         \bar{\Psi}(p,+)  + {\rm orth.}  &\quad
\psi_{p,\tau} = (-1)^{\tau} e^{(\epsilon-\mu)
\tau} \Psi(p,-)  + {\rm orth.}    \nonumber
\end{eqnarray}
It is interesting to note that the limit $a_{\rm T}
\rightarrow 0$ can be taken only {\em after} these
transformations when the $(-1)^{\tau}$
time-dependency mode has ``transmuted'' into an internal
degree of freedom.
The same kind of combinations have to be considered in the
spatial directions giving {\em their} flavour degeneracy.
However, as one is ultimately
interested in interacting fermions which
allow the continuum limit to be taken only at the very end
of the calculations, the alternating modes of free fermions
do not {\em a priori} signal any difficulties.

The baryon number as a path-integral observable
\cite{Kog83,HaK83} on the zero-time slice is
$$
B = \frac{\partial S}{\partial\mu} = \sum_p \left(
  e^{ \mu} \bar{\psi}_{p,-1} \psi_{p, 1}
+ e^{-\mu} \bar{\psi}_{p, 1} \psi_{p,-1}    \right)
$$
An intermediate result for the baryon number operator at
${\cal C}_0$ level is
$$
\hat{B} = \frac12 e^{\mu} \sum_{p:\,\epsilon>\mu} \left(
\frac1{\cosh\epsilon} \frac{\partial}{\partial\Psi(p,+)}
    + \bar{\Psi}(p,-) \right)
\left( e^{-\epsilon-\mu} \Psi(p,+)
+ \frac{e^{\epsilon-\mu}}{\cosh\epsilon}
  \frac{\partial}{\partial\bar{\Psi}(p,-)} \right) ~~~ +
$$ $$
- ~~~\frac12 e^{-\mu} \sum_{p:\,\epsilon>\mu} \left(
-\frac1{\cosh\epsilon}
\frac{\partial}{\partial\bar{\Psi}(p,-)}
    + \Psi(p,+) \right)
\left(- e^{\mu-\epsilon} \bar{\Psi}(p,-)
+ \frac{e^{\epsilon+\mu}}{\cosh\epsilon}
  \frac{\partial}{\partial\Psi(p,+)} \right)
$$ $$
+ ~~~\frac12 e^{\mu} \sum_{p:\, |\epsilon|<\mu} \left(
 \frac1{\cosh\epsilon} \frac{\partial}{\partial\Psi(p,+)}
-\frac1{\cosh\epsilon} \frac{\partial}{\partial\Psi(p,-)}
\right) \left(
e^{-\epsilon-\mu} \Psi(p,+) - e^{\epsilon-\mu} \Psi(p,-)
\right)
$$ $$
- ~~~\frac12 e^{-\mu} \sum_{p:\, |\epsilon|<\mu}
\left( \Psi(p,+) + \Psi(p,-) \right)
\left(
 \frac{e^{\epsilon+\mu}}{\cosh\epsilon}
 \frac{\partial}{\partial\Psi(p,+)}
+\frac{e^{\mu-\epsilon}}{\cosh\epsilon}
 \frac{\partial}{\partial\Psi(p,-)}
\right)
$$ $$
+ ~~~\frac12 e^{\mu} \sum_{p:\,\epsilon<-\mu} \left(
-\frac1{\cosh\epsilon} \frac{\partial}{\partial\Psi(p,-)}
    + \bar{\Psi}(p,+) \right)
\left(- e^{\epsilon-\mu} \Psi(p,-)
+ \frac{e^{-\epsilon-\mu}}{\cosh\epsilon}
  \frac{\partial}{\partial\bar{\Psi}(p,+)} \right)
$$ $$
- ~~~\frac12 e^{-\mu} \sum_{p:\,\epsilon<-\mu} \left(
\frac1{\cosh\epsilon}
\frac{\partial}{\partial\bar{\Psi}(p,+)}
    + \Psi(p,-) \right)
\left(e^{\mu+\epsilon} \bar{\Psi}(p,+)
+ \frac{e^{\mu-\epsilon}}{\cosh\epsilon}
  \frac{\partial}{\partial\Psi(p,-)} \right)
$$
This actually simplifies to
$$
\hat{B} = \sum_{p:\,\epsilon>\mu} \left(
  \bar{\Psi}(p,-) \frac{\partial}{\partial\bar{\Psi}(p,-)}
-  \Psi(p,+) \frac{\partial}{\partial\Psi(p,+)} \right)
+ \sum_{p:\,\epsilon<-\mu} \left(
  \bar{\Psi}(p,+) \frac{\partial}{\partial\bar{\Psi}(p,+)}
-  \Psi(p,-) \frac{\partial}{\partial\Psi(p,-)} \right)
$$
\begin{equation} \label{KSbaryon}
+ ~~~~ \sum_{p:\,|\epsilon|<\mu} \left( 1
-  \Psi(p,-) \frac{\partial}{\partial\Psi(p,-)}
-  \Psi(p,+) \frac{\partial}{\partial\Psi(p,+)} \right)
\end{equation}
The transfer matrix of Kogut-Susskind fermions is known
to be {\em not} of the preferable form $e^{-\hat{H}}$ with
$\hat{H}$ hermitian \cite{Sha81}. Nevertheless one can
define a quantum field theory using the two-step Hamiltonian
$ \hat{H} = -\frac12 \log T^2$. This is hermitian whenever
$\hat{T}$ is. Thus we obtain, generalizing to $\mu>0$,
$$
\hat{H} - \mu \hat{B} =
\sum_{p:\,\epsilon>\mu} \left( (\epsilon+\mu)
\Psi(p,+) \frac{\partial}{\partial\Psi(p,+)}
+ (\epsilon-\mu) \bar{\Psi}(p,-)
\frac{\partial}{\partial\bar{\Psi}(p,-)} \right)
$$ $$
+ \sum_{p:\, |\epsilon|<\mu} \left(
  (\epsilon+\mu) \Psi(p,+)
\frac{\partial}{\partial\Psi(p,+)}
+(-\epsilon+\mu) \Psi(p,-)
\frac{\partial}{\partial\Psi(p,-)}
\right)
$$ $$
+  \sum_{p:\,\epsilon<-\mu} \left( (-\epsilon+\mu)
\Psi(p,-) \frac{\partial}{\partial\Psi(p,-)}
+ (-\epsilon-\mu) \bar{\Psi}(p,+)
\frac{\partial}{\partial\bar{\Psi}(p,+)} \right)
$$
This can be rewritten in Fock form by appropriate
identification of baryon and antibaryon operators.
Let us put
$\epsilon_m={\rm arsinh}\,(\frac{a_{\rm T}}{a_{\rm S}}m)$
and $\eta=\pm$. Furthermore, to indicate
the relation between timelike ``flavour'' and the sign of
the spinorial energy, let $\eta p$ and $-\eta p$ stand
for $(p_1,p_2,p_3,\pm\sigma,s,f)$ and
 $(p_1,p_2,p_3,\mp\sigma,s,f)$, respectively.
The antibaryon annihilation and creation operators are
$$
\bar{b}(p,\eta) =
   \partial/\partial\Psi(\eta p, \eta) \qquad
\bar{b}^{\ast}(p,\eta) = \Psi(\eta p,\eta)
\qquad                        \epsilon(p)>\epsilon_m
$$
The corresponding baryon operators are
$$
b(p,\eta) =
 \partial/\partial\bar{\Psi}(-\eta p,\eta) \qquad
b^{\ast}(p,\eta) =
         \bar{\Psi}(-\eta p,\eta)
                       \qquad \epsilon(p)>\mu
$$ $$
b(p,\eta) =
 \Psi(-\eta p,\eta) \qquad
b^{\ast}(p,\eta) =
 \partial/\partial\Psi(-\eta p,\eta)
           \qquad \mu>\epsilon(p)>\epsilon_m
$$
We then have
$$
 \hat{B} = \sum_{p:\,\epsilon>\epsilon_m}
\sum_{\eta}
\left( b^{\ast}(p,\eta) b(p,\eta)
   -  \bar{b}^{\ast}(p,\eta)
        \bar{b}(p,\eta) \right)
$$
and
$$
\hat{H} - \mu \hat{B} = \sum_{p:\,\epsilon>\epsilon_m}
\sum_{\eta}
\left( (\epsilon-\mu) b^{\ast}(p,\eta) b(p,\eta)
   +   (\epsilon+\mu) \bar{b}^{\ast}(p,\eta)
        \bar{b}(p,\eta) \right)
+ \sum_{p:\,\mu>\epsilon>\epsilon_m} 2(\epsilon-\mu)
$$
We note that creation and annihilation operators are only
defined for multi-indices $p$ with a positive energy.
In terms of these operators the Hamiltonian and baryon
number take the standard form of baryons and antibaryons
with an extra two-component flavour $\eta$. The chemical
potential $\mu>0$ decreases the energy of baryons and
increases it for antibaryons, as expected.

As in the asymmetric case, the
$b^{\ast}$ and $\bar{b}^{\ast}$ at large $\epsilon$ or
$\mu$ (comparable to the {\sc uv} cutoff)
deviate from the hermitian adjoints of $b$ and $\bar{b}$:
$$
\bar{b}(p,\eta) = e^{-\epsilon-\mu}
   \cosh\epsilon ~ \bar{b}^{\ast}(p,\eta) \qquad
\epsilon>\epsilon_m
$$ $$
b^{\dag}(p,\eta) = e^{-\epsilon+\mu}\cosh\epsilon
     ~    b^{\ast}(p,\eta) \quad \epsilon>\mu
\qquad
b^{\dag}(p,\eta) =
 \frac{e^{-\epsilon+\mu}}{\cosh\epsilon} ~
 b^{\ast}(p,\eta) \quad \mu>\epsilon>\epsilon_m
$$
where $\epsilon=\epsilon(p)$.
The hermiticity of $\hat{H}$ is not affected by those
relations.

As with asymmetric fermions, the ground state functional is
$G_0[\Psi,\bar{\Psi}]=1$. In terms of creation and
annihilation operators this implies the same relations as
in (\ref{vac}) for both of the time-like flavours. The
ground state has a nonzero baryon density if
$\sinh\mu>\frac{a_{\rm T}}{a_{\rm S}}m$.
Thus, again, $\mu$ and $m$ are related in a nonlinear way,
reminding us of a degree of ambiguity in the
quantification of chemical potentials on the lattice
\cite{Bil84,Gav85}.

\section{Conclusions}

We have elaborated on how the sign of the ``energy'' of
Grassmann functional integral variables is associated with
the Fock-space degrees of freedom in the two cases of
Kogut-Susskind and time-asymmetric lattice fermions;
the central equations in this respect are (\ref{Psi}) and
(\ref{PsiKS}).
Our main conclusion is that the usual implementation of a
chemical potential for the baryon number, namely through
real-valued propagation factors in the discretized time
derivative for any kind of fermions, will indeed enhance or
deplete the baryons and antibaryons just as it is supposed
to do.

It is quite plausible that time evolution with
Kogut-Susskind fermions should involve some peculiarities.
In continuous Euclidean time, the positivity
of the transfer matrix is guaranteed already by the basic
requirement $\langle F|F\rangle\geq0$ \cite{OS73a,Lue77}.
By contrast, the Kogut-Susskind
one-step transfer matrix has both
positive and negative eigenvalues at {\em any}\/ time-like
lattice spacing \cite{Sha81} and is therefore usually
regarded as involving both time-evolution and flavour
rotation \cite{Smi84}.
In fact, the hermiticity
of the energy in this case is accomplished by
simply discarding the sign of the
negative eigenvalues of $\hat{T}$, namely in defining
$\hat{H}=-\frac12\log\hat{T}^2$ instead of
$\hat{H}=-\log\hat{T}$. The fact that this leads to a sound
quantum field theory in principle has been well-known
\cite{Sha81}. In the practice of Monte Carlo simulations,
when dealing with propagators of particles composed of
Kogut-Susskind fermions, a contribution of ``odd parity''
states (involving $(-1)^{\tau}$) is therefore fitted to a
free Kogut-Susskind particle with the same flavour
rotation and time evolution.

Complications arising from the nonpositivity of the transfer
matrix are avoided with Wilson fermions \cite{Lue77}, and
also with the asymmetric fermions of section 3, provided
that the temporal lattice spacing is sufficiently smaller
than the spatial one. For non-interacting fermions the
condition is $a_{\rm T}< a_{\rm S} / \sqrt{3+m^2}$;
similar bounds exist for fermions interacting with
a compact gauge field.

In computational practice, and related analytical study
such as \cite{HaK83,Bil84,Sta93},
the energy expectation value $\varepsilon$
of lattice fermions at finite temperature
is not derived from the logarithm of the transfer matrix,
$\varepsilon = {\rm tr}\,\hat{H} e^{-\hat{H}N_{\beta}}$,
but by differentiating the action with respect to the
temporal lattice spacing.
In terms of Fock space operators this can be obtained from
$\varepsilon = {\rm tr}\,\hat{E}\hat{T}^{N_{\beta}} =
{\rm tr}\,\hat{E} e^{-\hat{H}N_{\beta}}$
where $\hat{E}$ is the operator reconstructed from the
path-integral observable
$ \sum_p \bar{\psi}_{p,1} E(p) \psi_{p,1} $
for one-step fermions,
or $\frac12 \sum_p \left(
 \bar{\psi}_{p,-1} E(p) \psi_{p,-1}
+\bar{\psi}_{p,1} E(p) \psi_{p,1} \right)$
for Kogut-Susskind fermions. In fact, $\hat{E}$ and
$\hat{H}$ are different only for momenta comparable to the
{\sc uv} cutoff so that the use of $\hat{E}$ to obtain an
energy density is certainly justified by its computational
advantages.

It is reassuring to find that particles and
antiparticles are safely recognized by their internal
transformation properties (of which we have considered the
baryon number) and not by any kind of spinorial energy.
The latter would be an
ill-defined concept, indeed, in the presence of dynamical
gauge fields. The Hamiltonian of interacting fermions can
only be reconstructed as part of the complete,
translation-invariant fermion-boson field theory.
Naturally, such a reconstruction had to emphasize
local terms (as it did in \cite{Crz77,Lue77,Sha81})
rather than global diagonalizations.

As the time-asymmetric action
could provide an exact algorithm
with two flavours, it will be interesting to study
it further in the SU(3) interactive case, in concordance
with our view that the early onset problem could be due to
the excessive proliferation of degenerate flavours.

\subsection*{Acknowledgment}

We are grateful to U.-J. Wiese for an important hint to the
literature.


\newpage
\begin{thebibliography}{10}
\bibitem{Kog83} J. Kogut, H. Matsuoka, M. Stone, H. W. Wyld,
S. Shenker, J. Shigemitsu, D. K. Sinclair, Nucl. Phys. B 225
(1983) 93.
\bibitem{HaK83} P.Hasenfratz, F.Karsch, Phys. Lett. 125 B
(1983) 308.
\bibitem{Men93} E. Mendel, Nucl. Phys. B 30 (Proc.Suppl.)
(1993) 944.
\bibitem{Men92} E. Mendel, Nucl. Phys. B 387 (1992) 485.
\bibitem{Sta93} I.Bender, H.J.Rothe, W.Wetzel,
I.O.Stamatescu, Z. Phys. C 58 (1993) 333.
\bibitem{Sus77} L.Susskind, Phys. Rev. D 16 (1977) 3031.
\bibitem{Crz77} M.Creutz, Phys. Rev. D 15 (1977) 1128.
\bibitem{Lue77} M.L\"uscher, Commun. Math. Phys. 54 (1977)
283.
\bibitem{OSe78} K.Osterwalder, E.Seiler, Ann. Phys. 110
(1978) 440.
\bibitem{Sha81} H.S.Sharatchandra, H.J.Thun, P.Weisz, Nucl.
Phys. B 192 (1981) 205.
\bibitem{OS73a} K.Osterwalder, R.Schrader, Commun. Math.
Phys. 31 (1973) 33 and 42 (1975) 281; V.Glaser, Commun.
Math. Phys. 37 (1974) 257; J.Fr\"ohlich, K.Osterwalder,
E.Seiler, Ann. Math. 118 (1981) 461.
\bibitem{OS73b} K.Osterwalder, R.Schrader, Helv. Phys. Acta
46 (1973) 277.
\bibitem{Sei82} E.Seiler, Gauge theories as a problem of
constructive quantum field theory and statistical
mechanics, Lecture Notes in Physics 159, Springer 1982.
\bibitem{Sak85} B.Sakita, Quantum theory of many-variable
systems and fields, World Scientific 1985.
\bibitem{Smi84} M. Golterman, J. Smit, Nucl. Phys. B 245
(1984) 61.
\bibitem{Bil84} N. Bili\'c, R. V. Gavai, Z. Phys. C 23
                                                 (1984) 77.
\bibitem{Gav85} R. V. Gavai, Phys. Rev. D 32 (1985) 519.

\end{thebibliography}
\end{document}